\documentclass[rnote]{aa}  
\usepackage[active]{srcltx}
\usepackage{graphicx}
\usepackage{txfonts}
\usepackage{natbib}
\usepackage{multirow}
\bibpunct{(}{)}{;}{a}{}{,} % A&A style

\begin{document}

\title{Impact of the new solar abundances on the calibration of the PMS binary system RS Cha}
%   \subtitle{}

\author{
E. Alecian 
\inst{1}\and
Y. Lebreton 
\inst{2}\and
M-J. Goupil 
\inst{1}\and
M-A. Dupret
\inst{1}\and
C. Catala 
\inst{1}
}

\offprints{E.Alecian(evelyne.alecian@obspm.fr)}

\institute{
%1
Observatoire de Paris, LESIA, 5, place Jules Janssen, F-92195
Meudon Principal CEDEX, France \email{evelyne.alecian@obspm.fr}\and
%2
Observatoire de Paris, GEPI, UMR 8111, 5, place Jules Janssen, F-92195
Meudon Principal CEDEX, France
}

   \date{Received ; accepted }

% \abstract{}{}{}{}{} 
% 5 {} token are mandatory
 
  \abstract
  % context heading (optional), leave it empty if necessary  
	 {In a recent work, we tried to obtain a calibration of the two components of the pre-main sequence binary system RS Cha by means of theoretical stellar models. We found that the only way to reproduce the observational parameters of RS Cha with standard stellar models is to decrease the initial abundances of carbon and nitrogen derived from the GN93 solar mixture of heavy elements by a few tenths of dex .}
  % aims heading (mandatory)
   {In this work, we aim to reproduce the observational properties of the RS Cha stars with stellar evolution models based on the new AGS05 solar mixture recently derived from a three-dimensional solar model atmosphere. The AGS05 mixture is depleted in carbon, nitrogen and oxygen with respect to the GN93 mixture.}
  % methods heading (mandatory)
   {We calculated new stellar models of the RS Cha components using the AGS05 mixture and appropriate opacity tables. We sought models that simultaneously satisfy the observations of the two components (masses, radii, luminosities, effective temperatures and metallicity).}
  % results heading (mandatory)
   {We find that it is possible to reproduce the observational data of the RS Cha stars with AGS05 models based on standard input physics. From these models, the initial helium content of the system is $Y\sim0.255$ and its age is $\sim 9.13 \pm 0.12$ Myr.}
  % conclusions heading (optional), leave it empty if necessary 


   \maketitle

%
%________________________________________________________________

\section{Introduction}

Valuable tests of stellar physics are provided by the study of stars that are members of binary systems. When the global parameters of both components of such systems are determined accurately from observations, the calibration of the system by means of stellar evolution models may yield the fundamental properties of its members, like the age or initial helium content, as well as constraints on the input physics of the models.

We recently studied the double-lined eclipsing binary system RS Cha, the components of which are both on the pre-main sequence (PMS). We obtained spectroscopic data that allowed us to determine accurate values of the masses and radii of the components and the metallicity of the system \citep{alecian05}. We then complemented these parameters with previous determinations of the components' effective temperatures and luminosities  (Table \ref{tab:table1}) and we attempted to calibrate the system, i.e. we tried to fit the observational data by means of stellar evolution models that we calculated for the two stars for a common value of the age and initial chemical composition \citep[][hereafter paper I]{alecian06}. We showed that standard stellar models based on the \cite{grevesse93} solar mixture (hereafter GN93) are unable to satisfy the observational constraints for both stars even with reasonable changes in the model inputs. On the other hand, we found that both components can be fitted by standard stellar models of the same age and initial composition provided we decrease the C and N abundances of the GN93 solar mixture by a few tenths of dex.

Recently, the solar abundances have been revised by \cite{asplund04,asplund05a,asplund05b} on the basis of a time-dependent, three-dimensional hydrodynamical model of the solar atmosphere. In their analysis, \cite{asplund04,asplund05a,asplund05b} have considered departures from LTE and have used improved atomic and molecular data. The new solar mixture which is listed in \cite{asplund05b} (hereafter AGS05) is depleted in C, N, O with respect to the GN93 mixture. This in turn leads to a decrease in the solar $(Z/X)$ ratio (ratio of the mass of the heavy elements with respect to hydrogen).

In this paper, we present new stellar models based on the AGS05 solar mixture and appropriate opacity tables and we discuss their ability to reproduce the observational constraints of the two components of the RS Cha PMS binary system.

\begin{table}[htb]\def~{\hphantom{0}}
  \begin{center}
  \caption{Fundamental parameters of RS Cha. P stands for primary and S for secondary. References: 1: \citet{alecian05}, 2: \citet{ribas00}, 3: \citet{clausen80}}
  \label{tab:table1}
  \begin{tabular}{lccc}
  \hline
         & Primary   &   Secondary & References \\\hline%\\%[3pt]
             $M/M_\odot$ & $1.89\pm0.01$ & $1.87\pm0.01$ & 1 \\
             $R/R_\odot$ & $2.15\pm0.06$ & $2.36\pm0.06$ & 1 \\
$T_{\rm eff} {\rm (K)}$  & $7638\pm76  $ & $7228\pm72  $ & 2 \\
       $log L/L_\odot$   & $1.15\pm0.09$ & $1.13\pm0.09$ & $L=4\pi R^2 \sigma T_{\rm eff}^4$ \\
       $log\ {\rm g}$   & $4.05\pm0.06$ & $3.96\pm0.06$ & $g=G M/R^2$ \\
       $P (day)$ & \multicolumn{2}{c}{1.67}              & 1\\
       $i (degrees)$ &\multicolumn{2}{c}{$83.4\pm0.3$}   & 3 \\
       ${\rm [Fe/H]}$& \multicolumn{2}{c}{$0.17\pm0.01$} & 1 \\
  \hline
  \end{tabular}
  \end{center}
\end{table}

%
%__________________________________________________________________

\section{Physical ingredients and initial parameters of the stellar models}

We used the CESAM stellar evolution code \citep{morel97} to compute pre-main sequence stellar evolutionary tracks of constant mass in the range 1.86 to 1.90~$M_{\odot}$ corresponding to the range of masses of the RS Cha components (Table \ref{tab:table1}). We do not take into account the effects of rotation nor those of magnetic fields, and we neglect microscopic diffusion of the chemicals (see paper I). Each evolution is initialized with a homogeneous, fully convective model in quasi-static contraction and we define the age of a star as the time elapsed since initialization.

We calculated two series of models, each based on a different chemical mixture of heavy elements. Hereafter we refer to GN93 models, the models based on the \citet{grevesse93} solar mixture and to AGS05 models, the models based on the revised solar mixture obtained by \cite{asplund04,asplund05a,asplund05b}. The GN93 and AGS05 solar mixtures differ. The abundances of C, N, O of the AGS05 mixture are lower which in turn leads to a large decrease of the solar $(Z/X)$ ratio: $(Z/X)_{\odot}=0.0245$ for the GN93 mixture and $(Z/X)_{\odot}=0.0171$ for the AGS05 mixture.

Convection is treated using the classical mixing-length theory of convection \citep{bohm58}. The mixing-length parameter value is taken to be equal to $1.62 H_p$ ($H_p$ is the pressure scale-height). This value was obtained for the calibration of a solar model in luminosity and radius with the same input physics and the GN93 abundances. However the calibration of the solar model with the AGS05 mixture would change the solar mixing-length by less than $0.1 H_p$. Furthermore as both stars of RS Cha enclose a very small convective core and a radiative envelope, the change in the value of the mixing length parameter does not affect their position in the HR diagram.

We used the OPAL equation of state \citep{rogers96} and opacities calculated for the mixture used in the models. In the interior we used the OPAL opacities tables \citep{iglesias96} for either the GN93 or the AGS05 mixture (we generated these latter on the OPAL Web site) while at low temperatures (i.e. for $T\leq10^4$~K), we used either the \citet{alexander94} opacities -provided for the GN93 mixture- or the \cite{ferguson05} opacities available for the AGS05 mixture.

The species entering the nuclear reaction network are : $^1$H, $^3$He, $^4$He, $^{12}$C, $^{13}$C, $^{14}$N, $^{15}$N, $^{16}$O, $^{17}$O. We considered $^2$H, $^7$Li and $^7$Be to be in equilibrium. The nuclear reaction rates are taken from the NACRE compilation \citep{angulo99}.

%\begin{figure}
%\centering
%\includegraphics[width=5cm,angle=90]{opacity_temp.eps}
%\caption{Logarithm of the opacities plotted in function of the logarithm of the temperature for two different solar mixtures : Grevesse \& Noels (dashed line) and Asplund et al. (full line)}
%\label{fig:opa}
%\end{figure}

In our models, we take $Z/X=10^{\rm [Fe/H]}(Z/X)_{\odot}$, where $\rm [Fe/H]$ is the observed value given in Table \ref{tab:table1} and $(Z/X)_{\odot}$ is either the value of GN93 or the value of AGS05. We then derive the individual abundances of the heavy elements from the chosen solar mixture (GN93 or AGS05). 

We are aware that in \cite{alecian05} we obtained the metallicity of the system from a one dimensional ATLAS model atmosphere \citep{kurucz93} calculated with the GN93 old solar mixture. To be consistent, to calculate internal structure models based on the new solar mixture, we should have redetermined the $\rm [Fe/H]$-value of RS Cha from a three-dimensional model atmosphere calculated with the AGS05 mixture. However 3-D model atmospheres are still not available for A-stars which have finer convection zones and larger convection cells. On the other hand, it would be incorrect to use 1-D Kurucz models with the AGS05 abundance to redermine the metallicity of RS Cha.

\cite{alecian05} have determined the metallicity of RS Cha from measurements of the Fe and Ca abundances. Since the abundances of Fe and Ca are very close in the GN93 and AGS05 mixtures, it seems to be reasonable to assume that an analysis of the spectra with 3-D model atmospheres would not much change the $\rm[Fe/H]$-value of the system, which is a differential value. Under this assumption, we find that the revision of $(Z/X)_{\odot}$ from 0.0245 (GN93 mixture) to 0.0171 (AGS05 mixture) leads to a decrease of the $(Z/X)$ ratio of RS Cha from 0.0362 to 0.0253.

However for stars more massive than the Sun, work in progress seems to show that 1-D and 3-D model atmospheres yield non negligeable differences in $\rm [Fe/H]$ (Th\'evenin, private communication). To investigate the possibility that the 3-D determination leads to a value of $\rm [Fe/H]$  very different from the 1-D result of Table \ref{tab:table1}, we have calculated models with $\rm [Fe/H]$ values spanning the range 0.07-0.27 dex, that is $\pm 0.10$ dex around our mean value of $0.17$ dex.

The GN93 models presented below have been calculated with an initial helium mass fraction $Y_{\rm 0}=0.267$ derived from the calibration of the solar model in luminosity and radius. However we have shown in paper I that a change of the initial helium content does not change the behaviour of the GN93 models that cannot reproduce the observations. The calibration of the solar model in luminosity and radius of the solar model with the AGS05 abundances leads to an initial solar helium content that is lower by $\sim 0.01$ than that obtained with the GN93 abundances. In the new AGS05 models calculated for the present study, we have ajusted the initial helium content $Y$ to reproduce the observations, which yields $Y_0=0.255$ (see Sect. \ref{sec:3} below).

%In order to show that the calibration of RS Cha do , we calculated also models for two dfferent values of $\rm [Fe/H]$: 0.07 and 0.027.

%\begin{figure}
%\centering
%\includegraphics[height=5cm]{.ps}
%\caption{}
%\label{}
%\end{figure}

%
%______________________________________________________________

\section{Comparison with observations for GN93 and AGS05 stellar models}
\label{sec:3}

%In paper I we have not been able to reproduce the RS Cha observational constraints using standard models (first sample). We calculated new models using Asplund et al.'s mixture as well as the new opcity tables computed from new abundances (see Fig. \ref{fig:opa})(second sample).

\begin{figure}
\centering
\includegraphics[width=9cm]{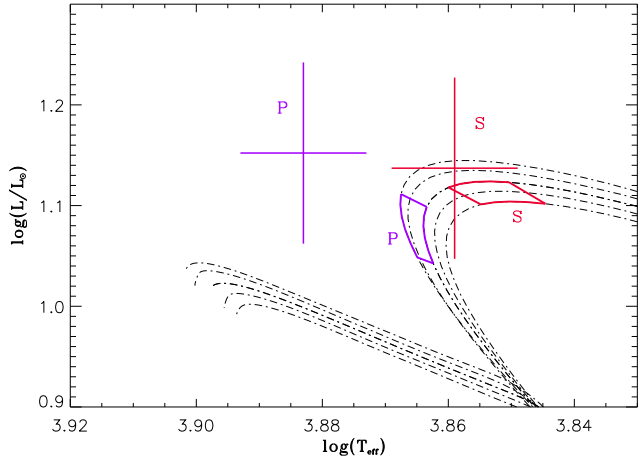}
\caption{Comparison of the GN93 models with observations. Crosses are the observed error bars in effective temperature and luminosity (Table \ref{tab:table1}) for the primary (P) component (left) and for the secondary (S) component (right). Dot-dashed lines are evolutionary tracks for the GN93 models of different masses: lowest to upper tracks are for masses of 1.86, 1.87, 1.88, 1.89 and 1.90 $M_{\odot}$. Boxes superimposed on the tracks with a full line correspond to models that satisfy the observed constraints on masses and radii.}
\label{fig2}
\end{figure}

\begin{figure}
\centering
\includegraphics[width=9cm]{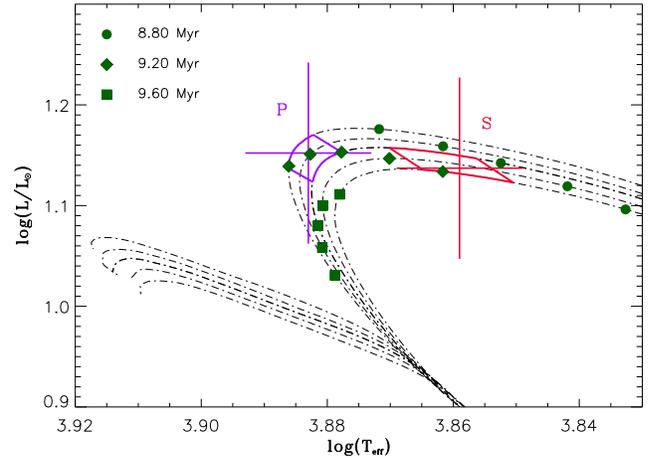}
\caption{Same as Fig. \ref{fig2} for the AGS05 stellar models. The dots, diamonds and squares indicate the ages of the models.}
\label{fig4}
\end{figure}

The RS Cha system is composed of two similar A-type stars in a PMS stage of evolution close to the ZAMS at the onset of CNO burning.
Figures \ref{fig2} and \ref{fig4} are H--R diagrams showing the comparison of the observations of the RS Cha components with stellar evolutionary tracks for the two mixtures considered. 

Observed effective temperatures and luminosities as well as masses and radii of the RS Cha components are compared to the corresponding values obtained from stellar model, as follows. For a given  mass $M_{\rm obs}$ given in Table \ref{tab:table1}, a stellar model is evolved until its radius matches the observed radius $R_{\rm obs}$. This provides the calculated luminosity and effective temperature $(L,T_{\rm eff})_{\rm calc}$ of the considered star. The same is done for the extreme values of the observed masses and radii and this gives rise to error boxes in effective temperature and luminosity which are superimposed onto the tracks in Figs. \ref{fig2} and \ref{fig4} with full lines. In the same graph, crosses represent the error bars in luminosity and effective temperature for the primary (left) and secondary (right) components derived from observations $(L,T_{\rm eff})_{\rm obs}$.

When both error boxes are located on their respective crosses, the corresponding stellar models reproduce the observations. As it can be seen in Fig. \ref{fig2}, this does not occur when GN93 models are used. On the other hand, Fig. \ref{fig4} shows that the AGS05 models calculated with $Y=0.255$ well match the observations. We point out that the AGS05 models can reproduce the observations without making any other change in the (standard) input physics of the models. 

As shown in paper~I the GN93 models cannot reproduce the observations, even if we consider reasonable uncertainties in the physical inputs or if we change their initial helium content and their metallicity within the error bars. The main problem is that  GN93 models predict that the luminosity of the primary ($L_{\rm P}$) is smaller than that of the secondary ($L_{\rm S}$) which is not supported by the observations. We also showed in paper~I that, in order to invert the luminosity ratio $L_{\rm P}/L_{\rm S}$, we need to delay the luminosity drop of the tracks toward the main sequence, induced by the onset of a convective core in the stars. This can only be obtained by decreasing the abundances of C and N. Indeed, as we remove carbon and nitrogen, the nuclear reactions involved in the CNO cycle become less efficient, and therefore the nuclear energy ($\epsilon$) produced in the center of the star decreases. We showed that the decrease of $\epsilon$ leads to a dominant decrease of the ratio $L/m$, $L$ and $m$ being the local luminosity and mass, compared to a negligible increase of the ratio $\frac{\kappa p}{T^4}$, where $\kappa$, $p$ and $T$ are the local opacity, pressure and temperature. The radiative gradient $\nabla_{\rm rad}$, proportional to $\frac{L}{m} \frac{\kappa P}{T^4}$, is therefore reduced, and the onset of the convective core is delayed. The AGS05 models therefore match the observations because the AGS05 mixture is depleted in C, N and O with respect to the GN93 mixture.
 
In order to ensure that the agreement would be kept even if the $\rm [Fe/H]$ value of RS Cha was to be modified by an analysis of the spectra based on three-dimensional model atmospheres, we calculated AGS05 models with $\rm [Fe/H]$-values of 0.07 and 0.27 dex.
We find that these models are still able to reproduce the observations after an adjustment of their initial helium abundance.

{The calibration of the RS Cha system based on the AGS05 stellar models and setting the same age for both components leads to an age of $9.13 \pm 0.12$ Myr for this PMS system, which is similar to the age we obtained in paper I from models where the C and N ratios had been decreased in the GN93 mixture. This age is the intersection of the age ranges given by both error boxes. The error bare is therefore very small, however it does not take into account the uncertainties of the physical input parameters such as the mixing length or the overshooting parameters. Recent studies show that changing these parameters in a reasonable range may lead to a modification of the age determination up to $\pm 20$ \% depending on the star considered \citep{lebreton04}. A reasonable error bar on our age determination would therefore be $\pm 2$ Myr.}

%
%______________________________________________________________

\section{Conclusions}

We close here a series of papers devoted to the observational and theoretical study of the two components of the PMS binary system RS Cha. In \cite{alecian05}, we revisited the fundamental parameters of the system and we determined its metallicity for the first time from spectroscopic observations. In a second paper \citep{alecian06}, we modeled the two stars of the system and have shown that stellar models built with standard input physics and the standard GN93 solar chemical mixture of heavy elements \citep{grevesse93} cannot reproduce the observations. We found that the only way to reproduce the observations is to decrease the C and N abundances by 0.6 and 0.5 dex respectively with respect to the GN93 mixture.

Here, we have obtained a satisfactory calibration of the RS Cha PMS binary system with stellar evolution models based on standard input physics and on the new AGS05 solar mixture recently derived by \cite{asplund04,asplund05a,asplund05b}. This is due to the fact that carbon, nitrogen and oxygen are depleted in the AGS05 solar mixture with respect to their values in the GN93 mixture.

To reproduce the observations, there are two alternatives:
\begin{itemize}
	\item The solar mixture of the heavy elements corresponds to the \cite{grevesse93} solar mixture. In this case, the abundance ratios of the chemical species in RS Cha cannot be solar and are typical of stars in young clusters \citep{daflon04}.
	\item The solar mixture of the heavy elements corresponds to the revised \cite{asplund04,asplund05a,asplund05b} solar mixture. In this case the abundance ratios of the species in RS Cha are compatible with a solar mixture.
\end{itemize}

As pointed out by \cite{asplund05b} the new AGS05 mixture corresponds to C and O abundances which are in better agreement with those measured in the local interstellar medium and nearby B stars. However they also make the agreement between the solar model and helioseismic observations much worse \citep[see for instance][]{basu04}. It would therefore be interesting to bring independent arguments in favor of (or against) the AGS05 mixture.

The available observations of RS Cha and of other members of the $\eta$ Cha young stellar cluster cannot favor one of the alternatives given above. Recent studies of the $\eta$ Cha cluster by \cite{lyo04,lyo06} identify 18 stars as members of the cluster, among which 15 are K-M type stars. A possibility to discriminate models calculated with the two mixtures would be to compare the corresponding isochrones with observations in the color-magnitude diagrams of  \cite{lyo04,lyo06}. {We checked that the isochrones calculated with CESAM for the GN93 mixture and a solar metallicity are close to those of \cite{siess00}. Furthermore we have estimated the age of the $\eta$ Cha cluster using RS Cha and the 5 more massive stars of the low-mass star sample of \cite{lyo04} (RECX 1, 4, 7, 10 and 11), for both mixtures : GN93 and AGS05. In both cases, we found a large age range consistent with the one found by \cite{lawson01}, from the \cite{dantona97,dantona98} tracks, and by \cite{lyo04} from the \cite{siess00} tracks. However for RECX 1, 7 and 10, being binaries \citep{lyo04}, we can reduce the age range by taking into account solely the single stars RECX 4 and 11. The age range obtained is consistent with our age determination of RS Cha whatever the mixture considered. Therefore we are not able to test the different solar mixtures with the low-mass stars of the $\eta$ Cha cluster with the isochrones fitting procedure.}

%However, we do not expect the 15 low-mass stars to be good candidates to test the different solar mixtures. At ages of $\sim10$ Myr they have not yet triggered nuclear reactions involving CNO elements and the isochrones will not show significant differences according to the solar mixture employed to calculate them. It remains only RS Cha and $\eta$ Cha to fit an isochrone which is not sufficient.

Another possibility would be to study the chemical composition of the $\eta$ Cha cluster by determining the carbon, nitrogen or oxygen abundances in the late-type stars of the cluster and to compare them to both solar mixtures. These abundances have not been determined yet, and spectroscopic observations of these stars would be helpful to support one of these alternatives.

In both cases, the models predict that the initial helium abundance of the system is subsolar and that its age is of about $9 \pm 2$ Myr, which is consistent with the recent age determination of \citet{lyo04,lyo06}.

\begin{acknowledgements}
We are very grateful to Fr\'ed\'eric Th\'evenin for fruitful discussions.
\end{acknowledgements}

%\nocite{*}
\bibliographystyle{aa}
\bibliography{rnrscha}

\end{document}